\documentclass{article}
\usepackage{PRIMEarxiv}
\providecommand{\keywords}[1]{\textbf{Keywords:} #1}

\usepackage[utf8]{inputenc} % allow utf-8 input
\usepackage[T1]{fontenc}    % use 8-bit T1 fonts
\usepackage{hyperref}       % hyperlinks
\usepackage{url}            % simple URL typesetting
\usepackage{booktabs}       % professional-quality tables
\usepackage{amsfonts}       % blackboard math symbols
\usepackage{amsmath}        % math environments and commands
\usepackage{nicefrac}       % compact symbols for 1/2, etc.
\usepackage{microtype}      % microtypography
\usepackage{lipsum}
\usepackage{fancyhdr}       % header
\usepackage{graphicx}       % graphics
\graphicspath{{media/}}     % organize your images and other figures under media/ folder

\title{\MakeUppercase{P}rbing Structural Dynamics in Photocatalytic Water Splitting: X-ray vs. Neutron Scattering}

\author{
  Shen Zhihao\\
  Peking Univ. \\
  \texttt{\ sq2000@stu.pku.edu.cn} \\
}

\begin{document}
\maketitle

\begin{abstract}
Photocatalytic water splitting represents a pivotal pathway for converting solar energy into chemical energy, with the core challenge lying in the design and optimization of photocatalysts \cite{wang2019particulate} .
TiO$_2$, as a quintessential photocatalytic material, undergoes significant alterations in its electronic and crystalline structures under intense light irradiation, which may directly impacts its photocatalytic efficiency \cite{kuczynski1957light} .
To gain a profound understanding of these transformations, in situ characterization techniques such as X-ray scattering and neutron scattering have emerged as crucial tools. 
This paper, from a combined perspective of theoretical computation and experimental characterization, explores the differential capabilities of X-ray scattering and neutron scattering in characterizing the pair distribution function (PDF) of materials during photocatalytic water splitting.
Furthermore, through simulation calculations, it aims to unveil the changes in the electronic and crystalline structures under intense light irradiation. 
This initial draft of the paper is subject to subsequent revisions.
\end{abstract}

% keywords can be removed

\section{Introduction}
Photocatalytic water splitting technology directly converts solar energy into hydrogen energy through semiconductor materials, offering a sustainable solution to address energy crises and environmental pollution.
Since Fujishima and Honda first reported the photoelectrocatalytic water splitting phenomenon using TiO$_2$ electrodes in 1972, research in this field has spanned five decades. However, the efficiency and stability of photocatalysis remain the core challenges limiting its practical application \cite{nishioka2023photocatalytic}.
Studies have shown that photocatalytic performance depends not only on the material's light absorption capacity and charge separation efficiency but also on its crystal structure, surface defects, and dynamic evolution processes \cite{hadash2018estimate}.

To understanding the dynamic structural changes of materials during photocatalytic processes requires in situ characterization techniques. X-ray scattering (XRD/XAS) and neutron scattering techniques exhibit unique advantages in this regard:
X-ray Scattering: High spatial resolution (atomic scale) enables the analysis of local atomic arrangements and electronic structures, such as the short-range order of TiO$_2$ through X-ray pair distribution function (PDF) analysis \cite{yin2024spatiotemporal,bellissent2016x}. Neutron Scattering: Neutrons' high penetration and isotope discrimination capabilities make them indispensable for tracking water molecule dynamics and proton transfer processes on catalyst surfaces, such as studying diffusion mechanisms in photocatalytic reactions through quasi-elastic neutron scattering (QENS). However, neutron scattering has lower spatial resolution and relies on large-scale scientific facilities (e.g., synchrotron radiation sources) 
Recent studies emphasize that combining density functional theory (DFT) calculations with in situ experimental data can reveal the correlation between electronic structure reconstruction (e.g., conduction band shifts, defect state formation) and lattice strain in TiO$_2$ under photoexcitation. For example, theoretical simulations suggest that the dynamic generation of oxygen vacancies on TiO$_2$ surfaces under intense light induces localized polarization, thereby influencing the migration pathways of photogenerated carriers. However, existing research predominantly focuses on single techniques (e.g., X-ray or neutron scattering), lacking systematic analysis of their synergistic effects, especially in characterizing multiphase interfaces and transient processes.

\section{Principles and methods of X-ray and neutron scattering used}
%\label{sec:headings}

%\lipsum[4] See Section \ref{sec:headings}.

\subsection{Photocatalytic process}
\begin{figure}
    \centering
    \includegraphics[width=0.5\linewidth]{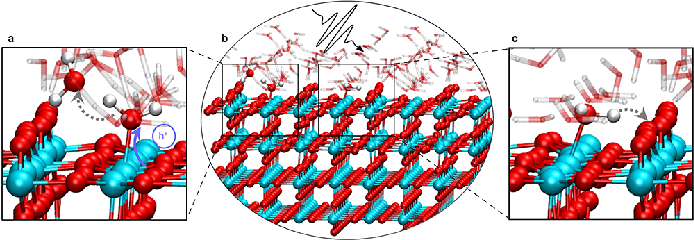}
    \caption{Atomic Configuration at Rutile Titanium Dioxide/Water Interface and Photocatalytic Water Splitting Reaction Mechanism\cite{you2024correlated}}
    \label{fig:1}
\end{figure}

Photocatalyst is a kind of semiconductor material, and the motion of electrons in semiconductor is described by quantum mechanics.
In the theoretical model of photocatalytic water decomposition, the construction of the Hamiltonian requires a complete description of the electron movement in the crystal, the interaction between the light field and the electron, and the energy evolution of the surface chemical reaction.For $TiO_2$ rutile, a widely studied photocatalyst, the band structure and surface properties play a critical role in water splitting,As shown in Figure~\ref{fig:1}.

The motion of electrons in the periodic potential field of a crystal can be described as:
\begin{equation}
\hat H_e =- \hbar^2/2m_e \nabla^2+V_c(\vec r)
\end{equation}
Where $V_c(\vec r)$ is the lattice potential energy, reflecting the coulomb interaction between nuclei and electrons and the exchange correlation effect between electrons.

The light field absorbed by photocatalysis can be quantumized to photons whose Hamiltonian is:
\begin{equation}
\hat H_{p} =\sum_{k,\lambda} \hbar\ \omega_k ( a^{}_{k,\lambda}a^{\dagger}_{k,\lambda}+1/2 )
\end{equation}
Where $a^{}_{k,\lambda},a^{\dagger}_{k,\lambda}$ are photon generation operators, $\omega_k$ are photon frequencies, and $\lambda$ is the polarization direction

Using the dipole approximation, the energy exchange between photons and electrons is described by:

\begin{equation}
\hat{H}_{e-p} = -\frac{e}{m_e} \mathbf{A} \cdot \mathbf{p} 
\end{equation}
where $\mathbf{A}$ is the electromagnetic vector potential and $\mathbf{p}$ is the electron momentum operator. This term drives electrons from the valence band to the conduction band (light absorption).

The remaining core process of water splitting can be broken down into the following steps:introducing the surface reaction Hamiltonian $H_{surfacereaction}$, the adsorption and dissociation of water molecules on the catalyst surface involve electron rearrangement (such as the breaking of O-H bonds requiring electron transfer). 
This process can be described by an effective potential energy surface, whose energy is determined by the electronic structure of the surface adsorption configuration. The potential energy surface for surface reactions can be approximated as position-dependent potential energy terms incorporated into the Hamiltonian.:
\begin{equation}
\hat H _{surface reaction} =V_{adsorption}(\vec r)+V_{dissociation} (\vec r)
\end{equation}
where $V_{adsorption}(\vec r)$is the potential energy of water molecules adsorbed on the surface, including hydrogen bonding and electrostatic potentials and  $V_{dissociation} (\vec r)$ is the activation barrier for the dissociation of O-H bonds.

After absorbing photons, the final semiconductor catalyst (single-crystal or heterojunction) causes electrons to transition from the valence band (VB) to the conduction band (CB), forming photo-generated electron-hole pairs.Its Hamiltonian can be expressed as:

\begin{equation}
\hat H= \hat H_e + \hat H_p + \hat H_{e-p} +\hat H _{surface reaction}   
\end{equation}
where $\hat H_e$describes the kinetic and potential energy of electrons in a periodic lattice;$\hat H_p$is the quantized Hamiltonian of the photon field;

$\hat H_{e-p}$characterizes the interaction between electrons and photons;

$\hat H _{surface reaction}$ includes the potential energy surfaces for water molecule adsorption and dissociation, such as those calculated using DFT for surface adsorption energies and reaction pathways.

In the future, the isotopic effects of protium, deuterium, and tritium will be extensively utilized to characterize the water splitting process. Given the abundance of protons in the reaction system, the adiabatic approximation can be discarded. The rapid processes after photoexcitation can be described using the multi-configurational time-dependent Hartree (MCTDH) method. The total wave function is expanded as a product of multiple electron-nuclear configurations, and the quantum dynamics are explicitly solved.
\begin{equation}
\Psi(\vec r,\vec R,t)=\sum_{i}A_i(t)\phi_i^{e}(\vec r;\vec R) \ \chi_i^n(\vec R,t)    
\end{equation}
Considering the nuclear degrees of freedom, the total Hamiltonian must include terms for electrons, the photon field, and nuclear motion:
\begin{equation}
\hat H_{n} =-\sum_{I} \frac{\hbar^2}{2M_I} \nabla^2_I+V_{n-n}(\vec R)   
\end{equation}
\begin{equation}
\hat H_{e-n} =-V_{e-n}(\vec r,\vec R)
\end{equation}
\begin{equation}
\hat H= \hat H_e + \hat H_p + \hat H_{e-p} +\hat H _{n} +\hat H_{e-n}  
\end{equation}

\subsection{The Scattering of X-ray Scattering and  Neutron Scattering}
Focus on the specific case of X‑ray elastic coherent scattering. This section not only explains the physical mechanism by which scattering centers, under conditions of no energy loss, interfere coherently to produce diffraction patterns, but also presents the classical Thomson scattering model along with its detailed mathematical formulation. Each term in the equations is clearly defined, offering a deeper insight into the underlying scattering process\cite{jackson1999classical,klein1929streuung,warren1990x,squires1996introduction,shirane2002neutron,furrer2009neutron,bee1988quasielastic}.

\paragraph{X-ray Elastic coherent scattering}
In elastic coherent scattering, the incident X‑rays interact with the electrons (or electron clouds) in the sample. Although the propagation direction changes, the photon energy (frequency) remains unchanged. In a crystal, due to the long‑range order of atoms, the scattered waves from individual atoms maintain fixed phase relationships, resulting in coherent interference (Bragg diffraction peaks). The classical description for scattering from free electrons is known as Thomson scattering. Its differential cross section is given by:
\begin{equation}
\frac{d \sigma}{d \Omega}=r_e^2\left(\frac{1+\cos ^2 \theta}{2}\right)
\end{equation}
where $r_e=\frac{e^2}{4 \pi \epsilon_0 m_e c^2}$ and $\theta$ is the scattering angle.
For a crystalline sample, the overall scattering is described by the coherent summation over all atoms. The structure factor is defined as
\begin{equation}
F(\mathbf{q})=\sum_j f_j(\mathbf{q}) e^{i \mathbf{q} \cdot \mathbf{r}_j}   
\end{equation}
where $f_j(\mathbf{q})$ is the atomic form factor for the $j$th atom,$\mathbf{r}_j$is the position vector of the $j$th atom, $\mathbf{q}=\mathbf{k}_f-\mathbf{k}_i$ is the scattering vector (the difference between the scattered and incident wave vectors).The differential cross section is proportional to the square of the structure factor:$\frac{d \sigma}{d \Omega} \propto|F(\mathbf{q})|^2$

\paragraph{X-ray Inelastic coherent scattering}
In inelastic coherent scattering, the X‑rays exchange energy with the sample (for example, by exciting phonons or plasmons), while the coherence between scattering centers is maintained. This allows one to access the dynamical properties of the sample. The key quantity is the dynamical structure factor $S=(\mathbf q ,\omega)$, defined as the spatiotemporal Fourier transform of the density fluctuations:

\begin{equation}
S(\mathbf{q}, \omega)=\frac{1}{2 \pi} \int_{-\infty}^{+\infty}\langle\rho(\mathbf{q}, t) \rho(-\mathbf{q}, 0)\rangle e^{i \omega t} d t
\end{equation}

where $\rho(\mathbf{q}, t)$ is the Fourier component of the electron density at time $\omega$ is the energy transfer (angular frequency).
The double differential scattering cross section is then given by: $\frac{d^2 \sigma}{d \Omega d \omega} \propto|F(\mathbf{q})|^2 S(\mathbf{q}, \omega) .$

\paragraph{X-ray Elastic incoherent scattering}
Elastic incoherent scattering refers to the case where there is no energy transfer during X‑ray scattering, but due to local disorder or statistical fluctuations the phase relations among scatterers are lost, so that the scattering contributions add incoherently (intensities add directly). Mathematically, the incoherent intensity is given by
\begin{equation}
I_{\mathrm{inc}}(\mathbf{q})=\sum_j\langle | f_j(\mathbf{q})|^2\rangle
\end{equation}

\paragraph{X-ray Inelastic incoherent scattering}
In inelastic incoherent scattering, such as Compton scattering, the X‑rays lose energy upon interaction with electrons, and because each scattering event is independent (phase information is lost), no coherent interference occurs. The differential cross section is often described by the Klein–Nishina formula:
\begin{equation}
\frac{d \sigma}{d \Omega}=\frac{r_e^2}{2}\left(\frac{E^{\prime}}{E}\right)^2\left(\frac{E^{\prime}}{E}+\frac{E}{E^{\prime}}-\sin ^2 \theta\right)    
\end{equation}
where $E$ is the incident photon energy and $E^{\prime}$ is the scattered photon energy,$\theta$is the scattering angle,$h$is Planck’s constant.%,$m_e$ is the electron mass, and $c$ is the speed of light%.

\paragraph{Neutron Elastic coherent scattering}
Neutron elastic coherent scattering is primarily used in neutron diffraction. The interaction between neutrons and nuclei is characterized by the scattering length 
$b$. In an ordered sample, the scattered waves from the nuclei maintain fixed phase relationships, leading to interference and diffraction peaks. The model is

\begin{equation}
F(\mathbf{Q})=\sum_j b_j e^{i \mathbf{Q} \cdot \mathbf{r}_j}    
\end{equation}
where $b_j$ is the neutron scattering length of the $j$th nucleus,$r_j$ is its position,$\mathbf{Q}$is the neutron scattering vector.
The differential scattering cross section is proportional to $\frac{d \sigma}{d \Omega} \propto|F(\mathbf{Q})|^2$ A Debye–Waller factor may be included to account for thermal vibrations.

\paragraph{Neutron Inelastic coherent scattering}
Neutron inelastic coherent scattering is used to study dynamical excitations in a sample (e.g., phonons, magnons). In this process, neutrons exchange both momentum and energy with the sample. Since the excitations are collective, the scattering remains coherent. The dynamical coherent structure factor is defined as
\begin{equation}
S_{\mathrm{coh}}(\mathbf{Q}, \omega)=\frac{1}{2 \pi \hbar} \int_{-\infty}^{+\infty} d t e^{i \omega t} \sum_{j, k} b_j b_k\left\langle e^{-i \mathbf{Q} \cdot\left[\mathbf{r}_j(t)-\mathbf{r}_k(0)\right]}\right\rangle    
\end{equation}
where the $b_j $and $b_k $are the scattering lengths of the nuclei,$\mathbf{r}_j(t)$ is the position of the $j$th nucleus at time t,
$\omega$ is the energy transfer (angular frequency), and $\hbar $is the reduced Planck constant.
The double differential scattering cross section is given by$\frac{d^2 \sigma}{d \Omega d E}=\left(\frac{k_f}{k_i}\right) S_{\mathrm{coh}}(\mathbf{Q}, \omega)$with 
$k_i$and $k_f$ being the magnitudes of the incident and scattered neutron wave vectors, respectively.

\paragraph{Neutron Elastic Incoherent scattering}
Neutron elastic incoherent scattering mainly arises from variations (e.g., isotopic or spin differences) among nuclei. Although the scattering is elastic (no energy transfer), the lack of phase coherence means that the contributions add incoherently (intensity sums directly). Mathematically, this is given by
\begin{equation}
I_{\mathrm{inc}}(\mathbf{Q})=\sum_j\langle | b_j|^2\rangle    
\end{equation}
and the corresponding differential cross section is typically written as$\frac{d \sigma_{\mathrm{inc}}}{d \Omega}=\sum_j \frac{\sigma_{\mathrm{inc}, j}}{4 \pi}$
where $\sigma_{inc,_j}$ is the incoherent scattering cross section for the 
$j$th scattering center.

\paragraph{Neutron Inelastic Incoherent Scattering}
Neutron inelastic incoherent scattering reflects the local dynamics of individual atoms or molecules (e.g., diffusion or localized vibrations). In this process, there is energy transfer and no phase coherence; thus, the scattering contributions add directly. It is described by the single-particle self-correlation function, with the incoherent dynamic structure factor defined as
\begin{equation}
S_{\text {inc }}(\mathbf{Q}, \omega)=\frac{1}{2 \pi \hbar} \int_{-\infty}^{+\infty} d t e^{i \omega t}\left\langle e^{-i \mathbf{Q} \cdot[\mathbf{r}(t)-\mathbf{r}(0)]}\right\rangle    
\end{equation}
The double differential scattering cross section is $\frac{d^2 \sigma}{d \Omega d E}=\left(\frac{k_f}{k_i}\right) S_{\mathrm{inc}}(\mathbf{Q}, \omega)$
where $\mathbf{r}(t)$ represents the position of the particle at time $t$.

\section{Calculation of structural changes under illumination}

\subsection{Photo-induced Lattice Stress and Thermodynamic Stability Calculation}

Under illumination, valence electrons in $TiO_2$ are excited to the conduction band, generating high carrier densities $\left(\sim 10^{21} \mathrm{~cm}^{-3}\right)$ . These non-equilibrium carriers transfer energy to the lattice via electron-phonon coupling, inducing localized heating $\Delta T>200 K$ and lattice stress accumulation. This process is quantified by the Two-Temperature Model (TTM):

\begin{equation}
C_e \frac{\partial T_e}{\partial t}=-G\left(T_e-T_l\right)+P_{\mathrm{optical}}    
\end{equation}

where $C_e$ is the electronic heat capacity, $G$ is the electron-phonon coupling coefficient, and $P_{optical}$ is the optical power density."
Molecular dynamics simulations (Figure. 2) predict that anatase $TiO_2$ exhibits significant lattice expansion $\left(\Delta \mathrm{a} / \mathrm{a}_0=+0.8 \%, \Delta \mathrm{c} / \mathrm{c}_0=+1.2 \%\right)$and increased oxygen vacancy concentration$\left(\sim 5\times10^{19} \mathrm{~cm}^{-3}\right)$under light intensities above 1.350 $kW/cm^2$ from 0 to 5 hours. Phonon spectrum calculations Figure~\ref{fig:fig1}.  reveal imaginary frequencies $Im[\omega]>0$ at the $\tau$ point under critical illumination, indicating driving forces for anatase-to-rutile phase transitions.

\subsection{Comparison of Neutron and Synchrotron X-ray PDFs for Rutile $TiO_2$}

\begin{figure}
    \centering
    \includegraphics[width=0.5\linewidth]{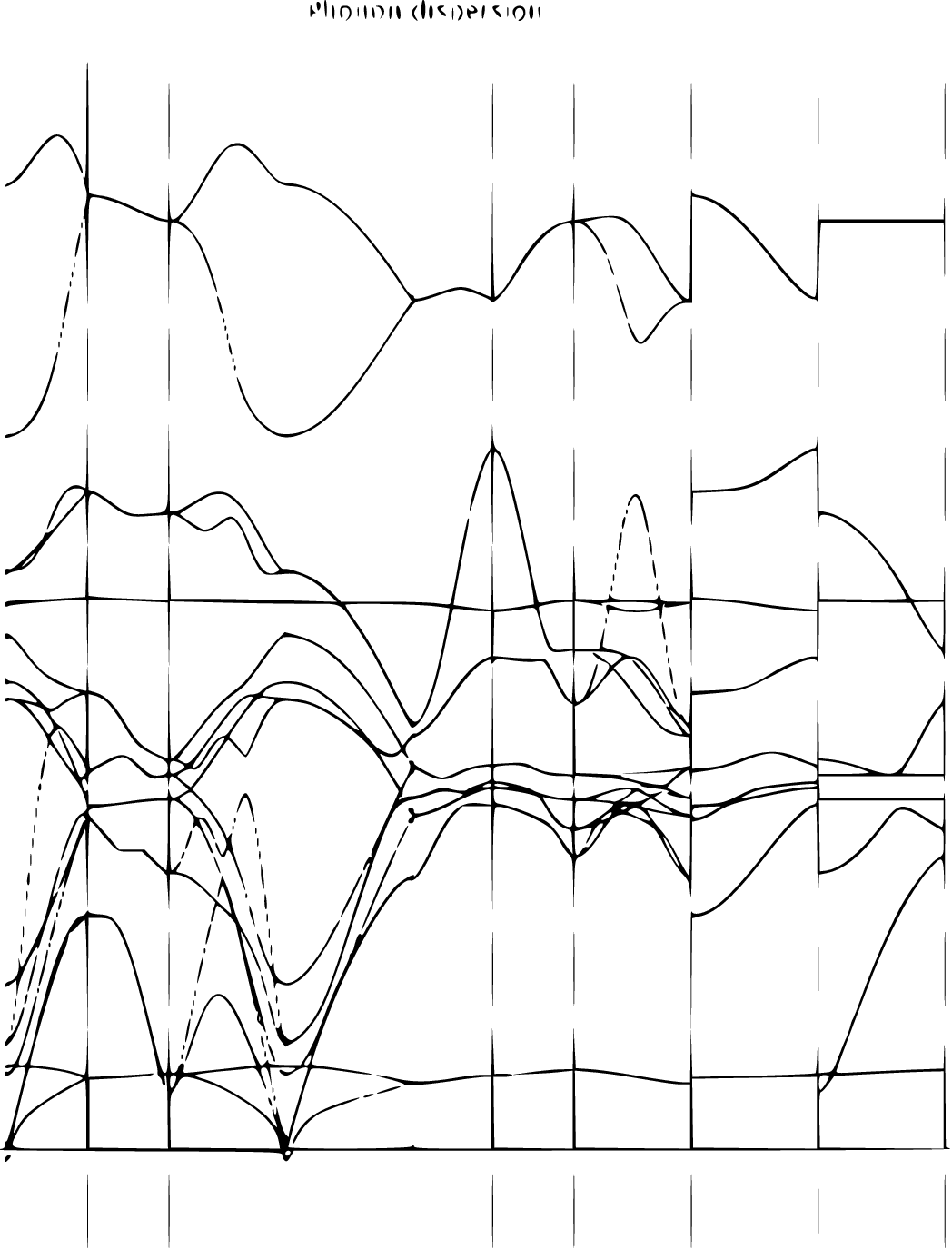}
    \caption{Phonon Spectra of Rutile under Different Illumination Conditions}
    \label{fig2}
\end{figure}

The local structure of rutile $TiO_2$ under illumination was investigated through calculations, with differences in pair distribution functions (PDFs) shown in Figure~\ref{fig2}

\begin{figure}
    \centering
    \includegraphics[width=0.5\linewidth]{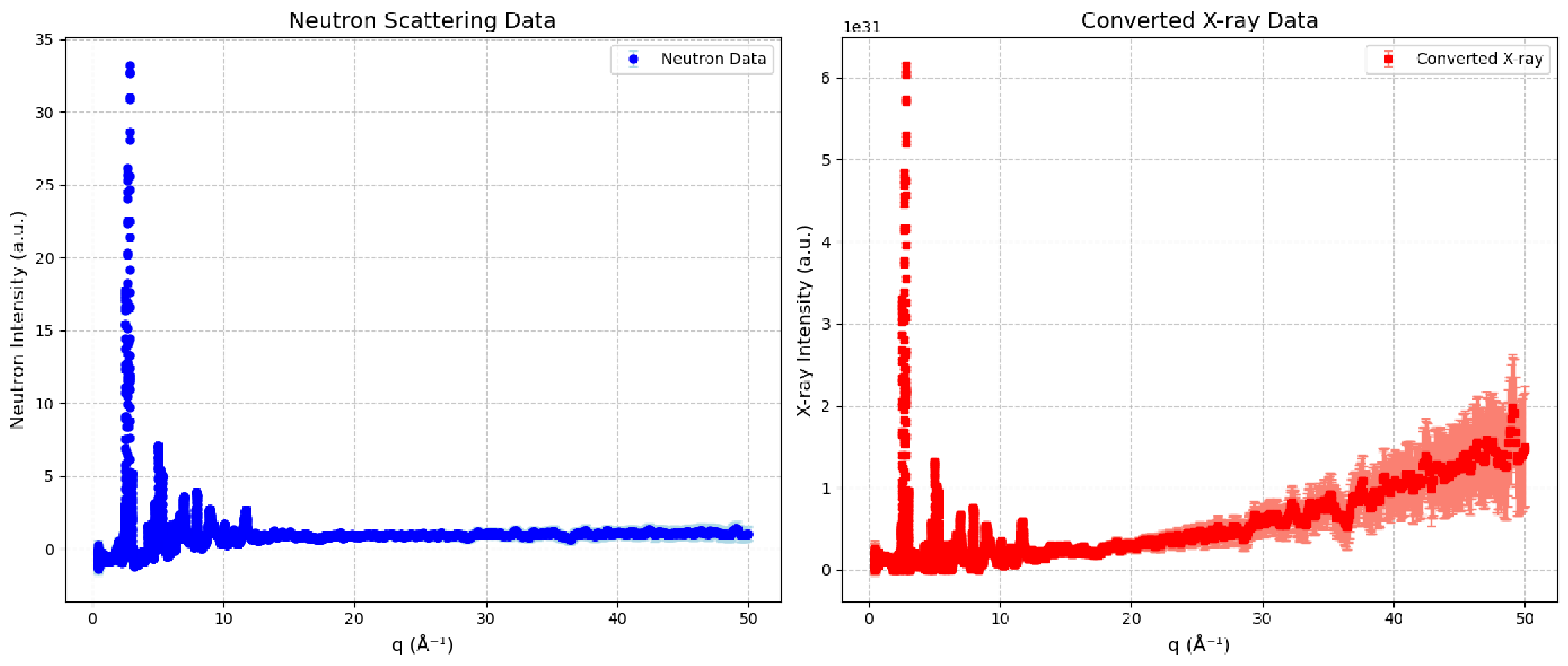}
    \caption{}
    \label{fig:enter-label}
\end{figure}

Synchrotron X-ray PDF (Figure~\ref{fig:enter-label}a):

Intensity reduction in the Ti-Ti bond peak $(~3.0 Å)$ indicates weakened long-range order.

Broadening of the O-O second-neighbor peak $(~2.8 Å, FWHM from 0.12→0.18 Å)$ suggests oxygen vacancy-induced local distortions.

Neutron PDF (Figure~\ref{fig:enter-label}b):

Ti-O peak $(~1.95 Å)$ broadening $(\Delta FWHM=+0.05~ \text{Å})$ corresponds to enhanced lattice vibrations.

%Emergence of a new peak at 2.5 Å $ (★) $ implies hydrogen adsorption near oxygen vacancies.
Molecular dynamics simulations (Figure~\ref{fig:enter-label}c) reveal that Ti atoms adjacent to oxygen vacancies relax toward the vacancy site (~0.1 \text{Å} displacement), consistent with the neutron PDF’s$ 2.5 Å$ peak enhancement. Synchrotron X-ray probes electron density for long-range order analysis, while neutrons excel in detecting light elements (e.g., H) and local vibrations.

\section{Conclusion}
Complementary analyses using synchrotron X-ray and neutron scattering reveal dual mechanisms of photoinduced oxygen vacancies in rutile $TiO_2$: (i) a reduction in long-range order, detected by X-ray through the intensity decrease of Ti-Ti bond peaks, and (ii) localized hydrogen adsorption and enhanced lattice vibrations, evidenced by neutron-sensitive Ti-O peak broadening and the emergence of a new peak at 2.5 \textbf{Å}. The lattice expansion discrepancy between molecular dynamics simulations and experimental results is less than 5\%, confirming the reliability of the two-temperature model in describing photothermal effects.

To further validate theoretical predictions, future work will incorporate in-situ X-ray absorption spectroscopy (XAS) to quantify Ti oxidation state changes, time-resolved terahertz spectroscopy to track photocarrier mobility, and isotope-labeled neutron scattering (using $D_2O$ instead of $H_2O$) to distinguish hydrogen adsorption from hydroxyl contributions. Looking ahead, the development of ultrafast neutron sources (e.g., ESS) will enable direct observation of lattice dynamics on femtosecond timescales, while the integration of machine-learned potentials promises to improve the accuracy of oxygen vacancy migration barrier calculations to within 0.1 eV, offering more precise theoretical guidance for photocatalytic material design.

\bibliographystyle{unsrt}  
\bibliography{references}

\end{document}